\documentclass[aps,pre,twocolumn]{revtex4}
\usepackage{graphicx}% Include figure files
\usepackage{dcolumn}% Align table columns on decimal point
\usepackage{bm}% bold math
\usepackage{epsfig}

\begin{document}

%\preprint{APS/123-QED}

\title{Electrophoresis of a polyelectrolyte through a nanopore}% Force line breaks with \\

\author{Sandip Ghosal}
 \email{s-ghosal@northwestern.edu}
\affiliation{%
Department of Mechanical Engineering, Northwestern University, Evanston, IL 60208
}%

%\affiliation{

\date{\today}% It is always \today, today,

\begin{abstract}
A hydrodynamic model for determining the electrophoretic speed of a polyelectrolyte 
through a nanopore is presented. It is assumed that the speed is determined 
by a balance of electrical and viscous forces arising from within the pore and that classical 
continuum electrostatics and hydrodynamics may be considered applicable. An explicit formula
for the translocation speed as a function of the pore geometry and other physical 
parameters is obtained and is shown to be consistent with experimental measurements 
 on DNA translocation through nanopores in silicon membranes. Experiments also show a 
 weak dependence of the translocation speed on polymer length that is not accounted 
 for by the present model. It is hypothesized that this is due to secondary effects that are
 neglected here. 
\end{abstract}

\maketitle

%\section{\label{sec:level1}First-level heading:\protect\\ The line
%break was forced \lowercase{via} \textbackslash\textbackslash}
%\section{Introduction}
Electrically driven translocation of DNA 
across natural and artificial nanopores
can be detected on the single molecule level by observing the increase of 
electrical resistance of the pore during such
events~\cite{kasianowicz_PNAS96}.
 Nanoscale pores may be fabricated  by the self assembly of the natural  protein 
$\alpha$-hemolysin on a lipid bilayer membrane~\cite{meller_etal_PNAS00,ssdna_sequence_nbt}
or by annealing a microfabricated hole in a $\mbox{Si/SiO}_{2}$ 
substrate using an intense electron beam from a TEM~\cite{storm_nature}.
The set up for detecting the translocation events consists of a reservoir containing an electrolyte 
that is partitioned into two chambers by a 
membrane with the nanometer scale pore forming the only 
communication between the two sides. An electrical potential difference 
 is applied across the membrane and the resulting 
current  is monitored. The passage 
of a DNA strand is signaled by a drop in the current. The duration 
and amount of these dips in the current contain signatures of the DNA
such as its length and base sequence. Possible applications of the technique to 
rapid DNA sequencing is being explored~\cite{ssdna_sequence_nbt}.

Lubensky and Nelson~\cite{lubensky_nelson}  provided an interpretation of some of the features 
of the experimental work cited above. 
 In particular, they  proposed a drift diffusion equation 
for $P(s,t)$: the probability that the DNA is found to have a length $s$ on a 
given side of the partition at time $t$. This equation was shown to explain qualitatively
the shape of the observed distribution of translocation times. It was presumed that the drift velocity itself is determined 
by a  resistive force that opposes the electrical driving force, but the exact
physical nature of this resistive force remained unclear. Viscous resistance could be 
a reasonable contender for a resistive force localized around the nanopore, but 
a simple  estimate appeared to indicate that its value was orders
of magnitude smaller than what was required. 

Polymer translocation across nanopores 
driven by a variety of physical mechanisms
have a number of other applications in biology;  the injection of 
DNA from a virus into a host cell is a particularly interesting 
example~\cite{muthukumar01}. These problems 
have been addressed by a number of 
authors ~\cite{sung_park,Boehm99,muthukumar99,muthukumar03,Kumar_Sebastian00,Lee_Sung_PRE01,sebastian_paul00}
by formulating the problem in a  probabilistic setting as a transition between two states separated by a 
barrier in the configurational entropy of the polymer chain. 
 In all cases, the hydrodynamic resistance of the pore when considered 
at all, is simply parametrized by a resistance coefficient. Most of the theoretical 
investigations on the subject to date are devoted to understanding how the translocation 
time of a polymer scales with polymer length, the scaling exponent being a quantity that may
 be expected to be independent of details such as the pore resistance and therefore 
 amenable to experimental verification.  
 
 In this paper, we explicitly determine the translocation time 
 by actually solving a simple hydrodynamic model for the 
translocation of a polymer through a water filled pore. Translocation speeds are calculated 
and compared  with the experimental data of Storm {\it et al.}~\cite{storm_physRevE05} for 
solid state nanopores. 

\begin{figure}
    \includegraphics[angle=0,width=2.5 in]{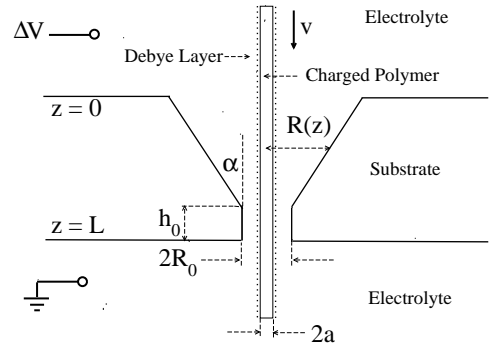}
    \caption{Translocation of a polyelectrolyte across a nanopore; geometry 
    of the pore region.}
    \label{geom}
\end{figure}

Figure~\ref{geom} is a sketch 
of the set up that also helps to explain our notation.
The pore is assumed to be cylindrically symmetric~\footnote{In the experiments reported 
in ~\cite{storm_physRevE05} the pore is actually pyramidal in shape tapering to a cone 
near the apex. However, since most of the resistance comes from the zone with the smallest gap 
width, the difference is not very significant.} about the $z$-axis and described 
by a function $R(z)$ that gives the distance to the wall of the pore from the axis at 
any $z$-location.
The part of the polyelectrolyte  within the pore is modeled as a straight cylinder 
of radius $a$ and carrying a uniform charge of density $\lambda$ per unit length. 
The length of the pore is $L$, 
and the  electric field, $E(z)$, is in the $z$-direction. 
Constancy of the current $I$ through any cross-section orthogonal to the z-axis requires:
\begin{equation}
E(z) = \frac{I}{\sigma \pi [R^{2}(z)-a^{2}]},
\label{ohms_micro}
\end{equation}
where $\sigma$ is the ionic conductivity of the buffer.
It is  assumed here that the electric field and current vectors make only 
a small angle to the $z$-axis, an assumption that is asymptotically approached 
in the limit $|R^{\prime}(z)| \ll 1$ for all $z$ where the prime indicates derivative.
The electrical conductivity $\sigma$ is taken as a constant in the pore 
with the same value as in the buffer. 

Next the fluid flow in the region between the polyelectrolyte cylinder and 
the walls of the nanopore is described, assuming that  classical continuum constant 
density hydrodynamics is applicable.
The assumption $|R^{\prime}(z)| \ll 1$ 
immediately suggests that the problem may be treated in the lubrication limit~\cite{batchelor,gh_02c},
so that (i) the pressure ($p$) is independent of the radial co-ordinate ($r$): $p=p(z)$
(ii) the flow ($\mathbf{u}$) is primarily along the $z$-axis, $\mathbf{u} \sim w(r,z) \hat{\mathbf{k}}$ 
and (iii) axial gradients are
negligible in comparison with radial ones. Therefore, $w$ satisfies
\begin{equation} 
- \frac{d p}{d z} + \frac{\mu}{r} \frac{d}{dr} \left( r \frac{d w}{d r} \right) = 0 
\label{stokes}
\end{equation} 
where $\mu$ is the dynamic viscosity of the buffer. Equation (\ref{stokes}) does not 
contain an electric body force term because, except for a thin Debye layer next to 
the polyelectrolyte, the bulk solution is electrically neutral.

In order to solve (\ref{stokes})  boundary conditions must now be specified. 
On the pore walls  the classical boundary condition of  {\em no slip}, is assumed:
\begin{equation}
w (R(z),z) = 0.
\label{noslip}
\end{equation}
The polyelectrolyte  cylinder itself will be surrounded by a charge cloud of counter-ions. 
The thickness of this Debye layer is measured by the Debye length $\lambda_{D}$,
which for the high salt ($\sim 1$ M KCl) buffers in these 
experiments  is extremely small: $\lambda_{D} \sim 0.3$ nm. 
The problem is therefore treated in the thin Debye layer limit, 
 which amounts to prescribing an apparent 
slip~\cite{helmholtz_1879,smoluchowski_1903} on charged surfaces (in SI units):
\begin{equation}
w(a,z) - v = - \frac{\epsilon \zeta E(z)}{\mu}.
\label{HSslip}
\end{equation} 
Here $v$ is the translocation velocity of the polyelectrolyte, $\zeta$ is the $\zeta$-potential 
at the surface of the polyelectrolyte and $\epsilon$ is the permittivity of the electrolyte.
The boundary condition (\ref{HSslip}) provides the coupling between the fluid 
and the electrical problems. 

The solution to (\ref{stokes}) with boundary conditions (\ref{noslip}) 
and (\ref{HSslip}) is
\begin{eqnarray} 
& & w(r,z) = - \frac{p^{\prime}(z)}{4 \mu} (R^{2} - r^{2} ) + \nonumber \\
& & \left\{ v + \frac{a^{2}}{R^{2}-a^{2}} \, u_e
+  \frac{p^{\prime}(z)}{4 \mu} (R^{2} - a^{2} ) \right\}\frac{\ln (R/r )}{\ln (R/a )},
\label{velocity}
\end{eqnarray}
where 
\begin{equation}
u_e = - \frac{\epsilon \zeta I}{\pi a^{2} \sigma \mu }
\label{defineue}
\end{equation}
defines a characteristic velocity scale for the problem. 
The pressure gradient $p^{\prime}(z)$ can be determined from the condition of
mass conservation,
\begin{equation} 
Q = \int_{a}^{R(z)} w(r,z) \, (2 \pi r) \, dr,
\label{flux} 
\end{equation}
where $Q$ is the flow rate through the pore. Substitution of (\ref{velocity}) in (\ref{flux}) 
gives 
\begin{eqnarray}
 \frac{a^{2}}{4 \mu} \frac{d p}{dz}  &=& \frac{Q}{\pi a^{2}} \frac{ 2\ln{R_{*}}}{(R_{*}^{2} -1)f} \nonumber\\
&  & - \left( v + \frac{u_e}{R_{*}^{2}-1} \right) \frac{R_{*}^{2} - 2 \ln{R_{*}} -1}{(R_{*}^{2}-1)f},
\label{pressure}
\end{eqnarray}
where $f = R_{*}^{2}-1 - (R_{*}^{2} + 1) \ln{R_{*}}$ and $R_{*} = R(z)/a$. 
The solution (\ref{velocity}) and (\ref{pressure}) still contains two undetermined parameters 
$v$ and $Q$. These are determined by imposing the conditions that there is no pressure 
difference across the pore,
\begin{equation}
p(0) = p (L)
\label{nopressure_drop}
\end{equation} 
and that the total force on the polyelectrolyte (which includes its Debye layer)  is zero
\begin{equation} 
\int_{0}^{L}  2 \pi a \mu  \left. \frac{\partial w}{\partial r}  \right|_{r=a} \, dz = 0. 
\label{noforce}
\end{equation}
The total force is zero since the polyelectrolyte moves through the pore without acceleration. 

Equations (\ref{nopressure_drop}) and (\ref{noforce}) yield after some algebra 
\begin{eqnarray} 
\frac{Q}{\pi a^{2} u_e}  &=& \frac{I_{0}I_{2} - I_{1} I_{3}}{I_{1} I_{2} + 2 I_{0} I_{2} -I_{0} I_{1}} \label{solnQ}\\
\frac{v}{u_e} &=& \frac{I_{0}^{2} - 2 I_{0} I_{3} - I_{0} I_{2}}{I_{1} I_{2} + 2 I_{0} I_{2} -I_{0} I_{1}} \label{solnv}
\end{eqnarray} 
where $I_0,I_1,I_2,I_3$ are dimensionless constants that depend solely 
on pore shape. They are defined as follows: 
$I_0 =  \langle f^{-1} \rangle$,   $I_1 =  \langle f^{-1} (R_{*}^{2}-1) \rangle$,
$I_{2} = \langle f^{-1} (R_{*}^{2}-1)^{-1} ( R_{*}^{2} - 2 \ln{R_{*}} -1 ) \rangle$ and 
$I_{3} = \langle f^{-1} (R_{*}^{2}-1)^{-2} ( R_{*}^{2} - 2 \ln{R_{*}} -1 ) \rangle$,
where $\langle \cdots \rangle = L^{-1} \int_{0}^{L} ( \cdots ) \; dz$ denotes average 
along the pore length.

\begin{table}[htdp]
\caption{Experimental parameters from Ref~\cite{storm_physRevE05} }
\begin{center}
\begin{tabular}{|c|c|c|c|c|c|c|c|}
\hline
$R_{0}$ & $a$  &  $L$ & $h_{0}$ & $\alpha$ & $\lambda_{D}$ & $\lambda$ & $\Delta V$\\ 
(nm)       &  (nm) &  (nm) & (nm) & (deg)      &    (nm)                  &  (e/nm)         & (mV)  \\      
\hline
5.0/4.0  &  1.0 & 340 & 40 & 36 & 0.3 & -5.9  & -120 \\
\hline
\end{tabular}
\end{center}
\label{table}
\end{table}

To calculate a numerical value for the translocation speed ($v$) from 
(\ref{solnv}), the $\zeta$-potential 
in (\ref{defineue}) must be related to the charge per unit length of the polyelectrolyte, $\lambda$
which is usually known from the chemistry. In order to do so, one must adopt 
some model for the structure of the Debye layer. For the purpose of this calculation, 
 the simplest model should suffice. Thus, it
is assumed that the Debye layer is described by the 
Poisson Boltzmann equation in  the 
Debye-Huckel limit~\cite{Russel}. Therefore, the electric potential due to the 
polyelectrolyte, $\phi(r)$ satisfies
\begin{equation} 
\frac{1}{r} \frac{d}{dr} \left( r \frac{d \phi}{d r} \right) = \frac{\phi}{\lambda_D^{2}}
\end{equation} 
with the boundary conditions 
$\phi(a) =\zeta$ and $\phi (\infty) = 0$.
The solution to this boundary value problem may be expressed in terms of the 
zero order modified Bessel function $K_{0}$:
\begin{equation} 
\phi (r) =   \frac{K_{0} (r / \lambda_D)}{K_{0} (a / \lambda_D) } \; \zeta.
\label{potential}
\end{equation} 
The linear charge density $\lambda$ can be related to the potential $\phi$ by 
Gauss's law:
\begin{equation} 
- 2 \pi a \, \phi^{\prime} (a) = \frac{\lambda}{\epsilon}.
\label{gauss}
\end{equation} 
Evaluating $\phi^{\prime}(a)$ from (\ref{potential}) and substituting in (\ref{gauss}) 
we have 
\begin{equation} 
\zeta = \frac{\lambda \lambda_{D} }{2 \pi a \epsilon} \,
\frac{K_0 (a / \lambda_D )}{ K_1 (a / \lambda_D ) },
\label{zetapot}
\end{equation} 
where $K_1$ is the modified Bessel function of order one. For the purpose of comparison 
with experimental data, it is convenient to replace the current $I$ in (\ref{defineue}) 
with the potential difference across the pore $\Delta V = V(0)-V(L)$, where $V(z)$ is the externally 
applied potential in the pore. Such a relation (Ohm's law) is readily obtained on 
integrating (\ref{ohms_micro}) between $z=0$ and $z=L$: 
$I/(\pi \sigma a^{2}) = \Delta V  / (I_{4} L)$ where $I_{4} = \langle (R_{*}^{2}-1)^{-1} \rangle$.
Therefore (\ref{defineue}) for $u_{e}$ may be written in an alternate form that does 
not involve the current or the conductivity,
\begin{equation} 
u_{e} = -  \frac{\epsilon \zeta}{\mu} \, \frac{\Delta V}{L} \, \frac{1}{I_{4}} = - \frac{u_{e0}}{I_{4}}.
\label{defineue0}
\end{equation} 
The quantity $u_{e0}=(\epsilon \zeta / \mu)( \Delta V/L)$ has a very simple
 interpretation, it is the velocity with which a particle 
of any shape and surface potential $\zeta$ will move if placed in an unbounded fluid medium 
and  acted upon by the average electric field that exists within the pore~\cite{morrison_70}. 
The funnel shape shown in Figure~\ref{geom} 
may be assumed for the solid state nanopores:
$R(z) = R_{0} + (L-h_{0}-z)  \tan \alpha$ if $z < L - h_{0}$
 and $R(z)=R_{0}$ otherwise,
$\alpha$ being the semi-vertical angle of the cone. The integrals $I_{0},I_{1},I_{2},I_{3}$ 
and $I_{4}$ are then evaluated numerically.

\begin{figure}
\includegraphics[width = 0.45\textwidth,angle=0]{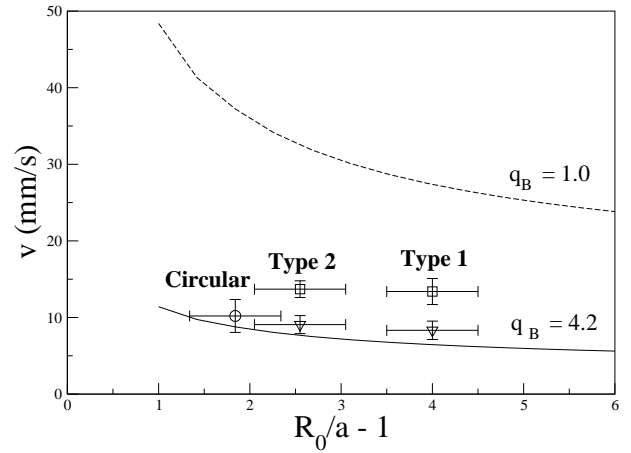}% Here is how to import EPS art
\caption{\label{figure} Translocation velocity ($v$)  as a function of the dimensionless 
gap width  ($R_{0}/a - 1$) in the case of 11.5 kbp  dsDNA (squares), 48.5 kbp 
dsDNA (triangles) and 11.5 kbp circular dsDNA (circle). Lines are the predictions from the 
model (a) assuming counterion condensation in accordance with the Manning theory for a line 
charge (solid) (b) ignoring counterion condensation (dashed).}
\end{figure}

The  prediction for the translocation speed $v$ given by (\ref{solnv}) is now compared 
with a set of measured values reported by 
Storm {\it et al.}~\cite{storm_physRevE05,storm_nature}. Table~\ref{table} summarizes 
the various parameters needed for such a comparison ($e$ is the charge on a proton).
The pore radius $R_{0}=4.0$ nm for the experiment with circular DNA (see Figure~\ref{figure}) 
but $5.0$ nm in all other cases.
The dielectric constant $\epsilon / \epsilon_{0} = 80$ and dynamic viscosity 
$\mu = 8.91 \times 10^{-4}$ Pa s for the electrolyte are taken as those of water.
Equation~(\ref{zetapot}) then gives~\footnote{A $\zeta$ potential 
also exists on the substrate-electrolyte interface, but at the very high counter-ion 
concentrations considered here ($\sim 1$ M) it's value is negligibly small compared 
to that of the polyelectrolyte~\cite{zetareview_eph04a}. }  $\zeta \approx - 56$ mV.
For highly charged polyelectrolytes such as DNA,
it has been shown by Manning~\cite{manning} and Oosawa~\cite{oosawa} that some of  the counterions  condense on to the surface of the polyelectrolyte reducing its effective 
charge to $\lambda_{\text{eff}}  =  \lambda/q_{B}$. For DNA, at 
room temperature, in weak salt solutions ($\lambda_{D} \gg a$) and away from boundaries, 
the Manning factor $q_{B} \approx 4.2$. However, under the conditions of the 
experiment, $\lambda_{D} \sim a$ and further, the polyelectrolyte is in the vicinity 
of a dielectric/conductor interface which has an effect on $q_{B}$ \cite{tang_jagota_hui_06}. 
Thus, the value of $\zeta$ is uncertain by perhaps as much as an order of magnitude.
The characteristic velocity $u_{e0}$ defined by (\ref{defineue0})  is
$u_{e0} \approx 15.7$ mm/s in the absence of counterion condensation  ($q_{B}=1$) but 
it is reduced to $3.7$ mm/s if $q_{B} = 4.2$ is assumed.

The lines in Figure~\ref{figure} show the translocation velocity $v$ calculated from 
(\ref{solnv}) as a function of $R_{0}/a - 1$.
The experimental data points are determined using Table I  and Figure~11 
of~\cite{storm_physRevE05}.
The translocation velocities are obtained by dividing the 
DNA length (11.5 kbp and 48.5 kbp) by the measured mean translocation times. The 
uncertainty in the velocity corresponds to the spread in the translocation times.
Four different modes can be distinguished~\cite{storm_physRevE05}
 for the DNA crossing, the two fundamental modes denoted as Type 1 and 2 and 
the mixed modes Type 12 and 21. In Type~1 
the DNA passes through the nanopore without any folds.  In the case of Type 2 
translocation of a linear chain or in the translocation of a circular chain, the DNA 
is folded in half so that its effective length is reduced to $L/2$ from $L$.
In that  case the two parallel DNA fibers that simultaneously thread 
the pore may be regarded as equivalent to a single cylindrical polyelectrolyte which has
the same $\zeta$-potential but a larger radius, $a_{\text{eff}}$. The effective radius 
is  determined by the requirement that 
the same fractional area of the pore is blocked. That is, the effective radius is 
$a_{\text{eff}} = \sqrt{2}  a \approx 1.41$ nm. Such a reduction has been used to calculate $R_{0}/a-1$ 
for the Type 2 and circular DNA cases
shown in Figure~\ref{figure}. An uncertainty of $\pm 0.5$ nm for the value of 
$R_{0}$ is assumed due to errors such as the departure of the nanopore from a 
strictly circular shape and  possible presence of hydration shells on the 
surface.

Lubensky and Nelson~\cite{lubensky_nelson} and Storm et al~\cite{storm_nanolett05} 
estimated the viscous force as $f_{v} \sim (2 \pi a L  \mu v )/(R_{0}-a) \sim 0.5$ -- $5.0$ pN,
if $v \sim 1-10$ mm/s. 
When compared with the electrical force $f_{el} \sim \lambda \,  \Delta V \sim 113$ pN,
it appeared that viscous friction could not balance the electrical traction and 
therefore other mechanisms were needed to explain the observed translocation rates. 
Figure~\ref{figure} demonstrates that this is not so. Two effects intervene to lower the 
estimate for the electrical force and raise the estimate for the viscous force 
(a) due to shielding by counter-ions the electrical force is reduced 
by as much as an order of magnitude~\cite{rabin_tanaka_05} 
(b) the change in 
the flow velocity takes place primarily across the thickness of a Debye layer, 
$\lambda_D$ and not over $R_{0}-a$, so that the actual viscous force is larger by a factor 
of approximately $(R_{0}-a)/\lambda_D  \sim 10$. In fact, in the thin Debye layer 
limit, the  solution for the fluid flow within the Debye layer guarantees an exact 
cancellation of the electrical force with the  viscous force on every surface 
element $dS$ of the polyelectrolyte. 
Figure~\ref{figure} also shows that the translocation velocity does have a weak dependence 
on polymer length. In a certain range of polymer lengths, the data 
can be fit by a power law: $v \sim L^{-0.26}$~\cite{storm_nanolett05}. This fact  cannot 
be explained by the model presented here or by any other model 
that localizes all of the resistive force within the pore. The dependence on polymer 
 length must arise from additional effects not considered here such as the 
 viscous  resistance from the part of the polymer that lies outside 
the pore~\cite{storm_nanolett05}.

The largest source of uncertainty in the above calculation arises from the difficulty 
of obtaining accurately a value of $\zeta$. The current understanding of the physics 
of the Debye layer is still incomplete so that even in the classical 
problems of electrokinetics, such as in the electrophoresis of a sphere, unresolved 
discrepancies exist between (\ref{HSslip}) and the $\zeta$ determined from 
more direct measurements of charge~\cite{kim_netz_JCP_2006}.
The use of the lubrication equations and the assumption of a simplified axi-symmetric 
geometry for the pore also contribute to the error but 
these are likely to be much less than the ones just mentioned. 
One may question the use of continuum hydrodynamics in the first place 
to calculate the mean motion of the polymer. Note, however,
in the case of  water, the intermolecular spacing is of the order 
of $0.1$ nm which is significantly smaller than the $\sim 2$ nm inner diameter 
and $4-5$ nm outer diameter of the pore. 
 The applicability of the no slip condition at the
solid liquid  interfaces is still  a matter of contention~\cite{lauga_brenner_stone_noslip06}.
Nevertheless, possible slip lengths are miniscule and amounts to an
uncertainty in the values of $R_{0}$ and $a$  by perhaps a fraction of a nanometer.
In fact, classical hydrodynamics (i.e. Stokes equations with no slip boundary conditions)
work reasonably well for water down to several tenths of nanometers; for example, 
for non-electrolytes, molecular sizes calculated on the basis of the Stokes-Einstein relation or Einstein's 
viscosity law for dilute suspensions agree with molecular structure based determinations 
to within $10-15$ percent~\cite{schultz_solomon_JGenPhys61} even for molecules 
in the $0.3$--$0.5$ nm range. The use of the continuum hydrodynamic model for 
calculating statistical averages in the manner used here is therefore not likely 
to be a significant source of uncertainty, at least for the solid state nanopores.

In summary, a model of the pore resistance based on continuum hydrodynamics 
and electrostatics produces estimates for the translocation speed of dsDNA in 
solid state nanopores  to within an order of magnitude of experimental values. 
The present analysis needs to be modified for  protein nanopores 
because certain approximations such as  the thin Debye layer 
are not applicable in that case.

\noindent {\em Acknowledgement:} The author wishes to thank Prof. David Deamer for helpful discussions.

%\bibliographystyle{/Users/sgh219/Documents/Work/LIBRARY/BIBFILES/prsty}
%\bibliography{/Users/sgh219/Documents/Work/LIBRARY/BIBFILES/membrane,/Users/sgh219/Documents/Work/LIBRARY/BIBFILES/microfluidics}

\end{document}